\documentclass[
  abstract=true,
  oneside,
  listof=totoc,
  numbers=noenddot,
  bibliography=totoc,
  headsepline,
  final,
  english
]{scrartcl} 

\usepackage[english]{babel}
\usepackage[english]{isodate}

\usepackage{amsmath} 

\usepackage[
  backend=biber,
  style=ieee,
  alldates=long
]{biblatex}

\usepackage{csquotes}
\usepackage{graphicx}           
\usepackage{caption}            
\usepackage{subcaption}         
\usepackage{listings}           
\usepackage[table]{xcolor}
\usepackage{booktabs}           
\usepackage{microtype}          
\usepackage{units}              
\usepackage{array}
\usepackage{fancybox}           
\usepackage{units}              
\usepackage{fancyvrb}           
\usepackage{pdfpages}
\usepackage{hyphenat}
\usepackage{todonotes}
\usepackage{xspace}
\usepackage{setspace}

\usepackage{microtype}

\usepackage{mathtools}

\usepackage{dsfont}

\usepackage{float}

\usepackage{amssymb}            
\usepackage{scrhack}

\usepackage[
  citebordercolor={0.75 0.75 1},
  filebordercolor={0.75 0.75 1},
  linkbordercolor={0.75 0.75 1},
  urlbordercolor={0.75 0.75 1},
  pdfborder={0.75 0.75 1},
  plainpages=false,
  pdfpagelabels=true
]{hyperref}

\usepackage{cleveref}
\crefname{subsection}{subsection}{subsections}

\vfuzz \hfuzz
\def \FigWidth {0.8\textwidth}

\definecolor{mygreen}{rgb}{0,0.6,0}
\definecolor{mygray}{rgb}{0.5,0.5,0.5}
\definecolor{mymauve}{rgb}{0.58,0,0.82}

\newcommand{\quotes}[1]{``#1''} 

\makeatletter
\AtBeginDocument{
  \hypersetup{
    pdftitle = {multivar_horner},
    pdfauthor = {Jannik Michelfeit}
  }
}
\makeatother

\begin{document}
 \pagenumbering{arabic}

\title{multivar\_horner: a python package for computing Horner factorisations of multivariate polynomials}

\author{
Jannik Michelfeit\\
Technische Universität Dresden\\
\href{mailto:jannik@michelfe.it}{\texttt{jannik@michelfe.it}}
}

\date{\today}

\maketitle

\textbf{keywords:} multivariate polynomials, Horner factorisations, Horner factorizations, \\
polynomial evaluation, python package, open source software

\begin{abstract}
Many applications in the sciences require numerically stable and computationally efficient evaluation of multivariate polynomials.
Finding beneficial representations of polynomials, such as Horner factorisations, is therefore crucial.
\mbox{\texttt{multivar\_horner}}\cite{multivarhorner}, the \mbox{\texttt{python}} package presented here, is the first open source software for computing multivariate Horner factorisations.
This work briefly outlines the functionality of the package and puts it into reference to previous work in the field.
Benchmarks additionally prove the advantages of the implementation and Horner factorisations in general.
\end{abstract}

\section{Introduction}

Polynomials are a central concept in mathematics and find application in a wide range of fields.
(Multivariate) polynomials have different possible mathematical representations and the beneficial properties of some representations are in great demand in many applications\cite{LeeFactorization2013, leiserson2010efficient, Hecht1}.

The so called Horner factorisation is such a representation with beneficial properties.
Compared to the unfactorised representation of a multivariate polynomial, in the following called \emph{canonical form}, this representation offers some important advantages.
First of all the Horner factorisation is more compact in the sense that it requires less mathematical operations in order to evaluate the polynomial (cf. \cref{fig:num_ops_growth}).
Consequently, evaluating a multivariate polynomial in Horner factorisation is faster and numerically more stable\cite{pena2000multivariate, pena2000multivariate2, greedyHorner} (cf. \cref{fig:num_err_growth}).
These advantages come at the cost of an initial computational effort required to find the factorisation.

The \mbox{\texttt{multivar\_horner}} \mbox{\texttt{python}} package implements a multivariate Horner scheme (\quotes{Horner's method}, \quotes{Horner's rule})\cite{horner1819xxi} and thereby allows computing Horner factorisations of multivariate polynomials given in canonical form.
Representing multivariate polynomials of arbitrary degree also in canonical form, computing derivatives of polynomials and evaluating polynomials at a given point are further features of the package.
Accordingly the package presented here can be helpful always when (multivariate) polynomials have to be evaluated efficiently, the numerical error of the polynomial evaluation has to be small or a compact representation of the polynomial is required.
This holds true for many applications applying numerical analysis.
One example use case where this package is already being employed are novel response surface methods\cite{michelfeitresponse} based on multivariate Netwon interploation\cite{Hecht1}.

\section{Functionality}

In its core \mbox{\texttt{multivar\_horner}} implements a multivariate Horner scheme with the greedy heuristic presented in \cite{greedyHorner}.
In the following the key functionality of this package is being outlined.
For a more details on polynomials and Horner factorisations please refer to the literature, e.g. \cite{neumaier2001introduction}.

A polynomial in canonical form is a sum of monomials.
For a univariate polynomial, which can be written as $f(x) = a_0 + a_1 x + a_2 x^2 + ... + a_d x^d$ (canonical form), the Horner factorisation is unique: $f(x) = a_0 + x ( a_1 + x( ... x (a_d) ... )$
In the multivariate case however the factorisation is ambiguous, as there are multiple possible factors to factorise with.
The key functionality of \mbox{\texttt{multivar\_horner}} is finding a good instance among the many possible Horner factorisations of a multivariate polynomial.

Let's consider the example multivariate polynomial in canonical form $p(x) = 5 + 1 x_1^3 x_2^1 + 2 x_1^2 x_3^1 + 3 x_1^1 x_2^1 x_3^1$.
The polynomial $p$ is the sum of $5$ monomials, has dimensionality $3$ and can also be written as $p(x) = 5 x_1^0 x_2^0 x_3^0 + 1 x_1^3 x_2^1 x_3^0 + 2 x_1^2 x_2^0 x_3^1 + 3 x_1^1 x_2^1 x_3^1$.
The coefficients of the monomials are $5$, $1$, $2$ and $3$ respectively.
It is trivial but computationally expensive to represent this kind of formulation with matrices and vectors and to evaluate it in this way.
In this particular case for example a polynomial evaluation would require 27 operations.
Due to its simplicity this kind of formulation is being used for defining multivariate polynomials as input.
The following code snipped shows how to use \mbox{\texttt{multivar\_horner}} for computing a Horner factorisation of $p$ and evaluating $p$ at a point $x$:

\begin{lstlisting}[language=python]
from multivar_horner import HornerMultivarPolynomial
coefficients = [5.0, 1.0, 2.0, 3.0]
exponents = [[0, 0, 0], [3, 1, 0], [2, 0, 1], [1, 1, 1]]
p = HornerMultivarPolynomial(coefficients, exponents, rectify_input=True)
# [#ops=10] p(x)=x_1(x_1(x_1(1.0 x_2)+2.0 x_3)+3.0 x_2 x_3)+5.0
x = [-2.0, 3.0, 1.0]
p_x = p.eval(x, rectify_input=True) # -29.0
\end{lstlisting}

The factorisation computed by \mbox{\texttt{multivar\_horner}} is $p(x) =  x_1 (x_1 (x_1 (1 x_2) + 2 x_3) + 3 x_2 x_3) + 5$ and requires 10 operations for every polynomial evaluation.
It should be noted that the implemented factorisation procedure is coefficient agnostic and hence does not for example optimise multiplications with $1$. 
This design choice has been made in order to have the ability to change the coefficients of a computed polynomial representation a posteriori.

With the default settings a Horner factorisation is being computed by recursively factorising with respect to the factor most commonly used in all monomials.
When no leaves of the resulting binary \quotes{Horner Factorisation Tree} can be factorised any more, a \quotes{recipe} for evaluating the polynomial is being compiled.
This recipe encodes all operations required to evaluate the polynomial in \mbox{\texttt{numpy}} arrays\cite{numpy}.
This enables the use of functions just in time compiled by \mbox{\texttt{Numba}}\cite{numba}, which cause the polynomial evaluation to be computationally efficient.
The just in time compiled functions are always being used, since polynomial evaluation in pure \mbox{\texttt{python}} would to some extent outweigh the benefits of Horner factorisation representations.

\section{Degrees of multivariate polynomials}

It is important to note that in contrast to the one dimensional case, several concepts of degree exist for polynomials in multiple dimensions.
Following the notation of \cite{trefethen2017multivariate} the usual notion of degree of a polynomial, the maximal degree, is the maximal sum of exponents of all monomials.
This is equal to the maximal $l_1$-norm of all exponent vectors of the monomials.
Accordingly the euclidean degree is the maximal $l_2$-norm and the maximal degree is the maximal $l_{\infty}$-norm of all exponent vectors.
Refer to \cite{trefethen2017multivariate} for precise definitions.

A polynomial is called fully occupied with respect to a certain degree if all possible monomials having a smaller or equal degree are present.
The occupancy of a polynomial can then be defined as the amount of existing monomials relative to the fully occupied polynomial of this degree.
A fully occupied polynomial hence has an occupancy of $1$.

\begin{figure}[tbh]
\centering
\includegraphics[width=\FigWidth]{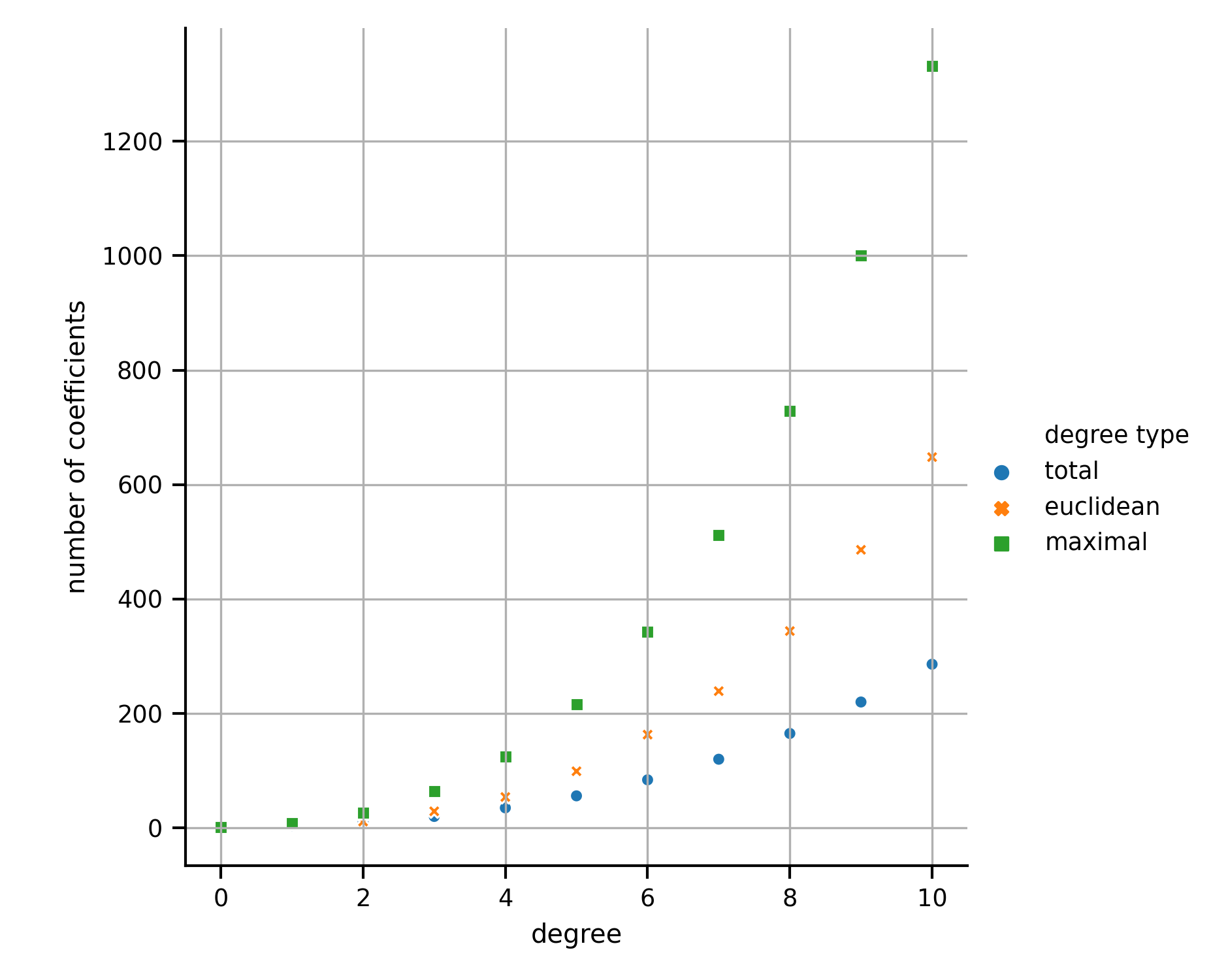}
\caption{the amount of coefficients of fully occupied polynomials of different degrees in 3 dimensions.}
\label{fig:num_coeff_growth}
\end{figure}
\noindent

The amount of coefficients (equal to the amount of possible monomials) in multiple dimensions highly depends on the type of degree a polynomial has (cf. \cref{fig:num_coeff_growth}).
This effect intensifies as the dimensionality grows.

\section{Benchmarks}

\begin{figure}[tbh]
\centering
\includegraphics[width=\FigWidth]{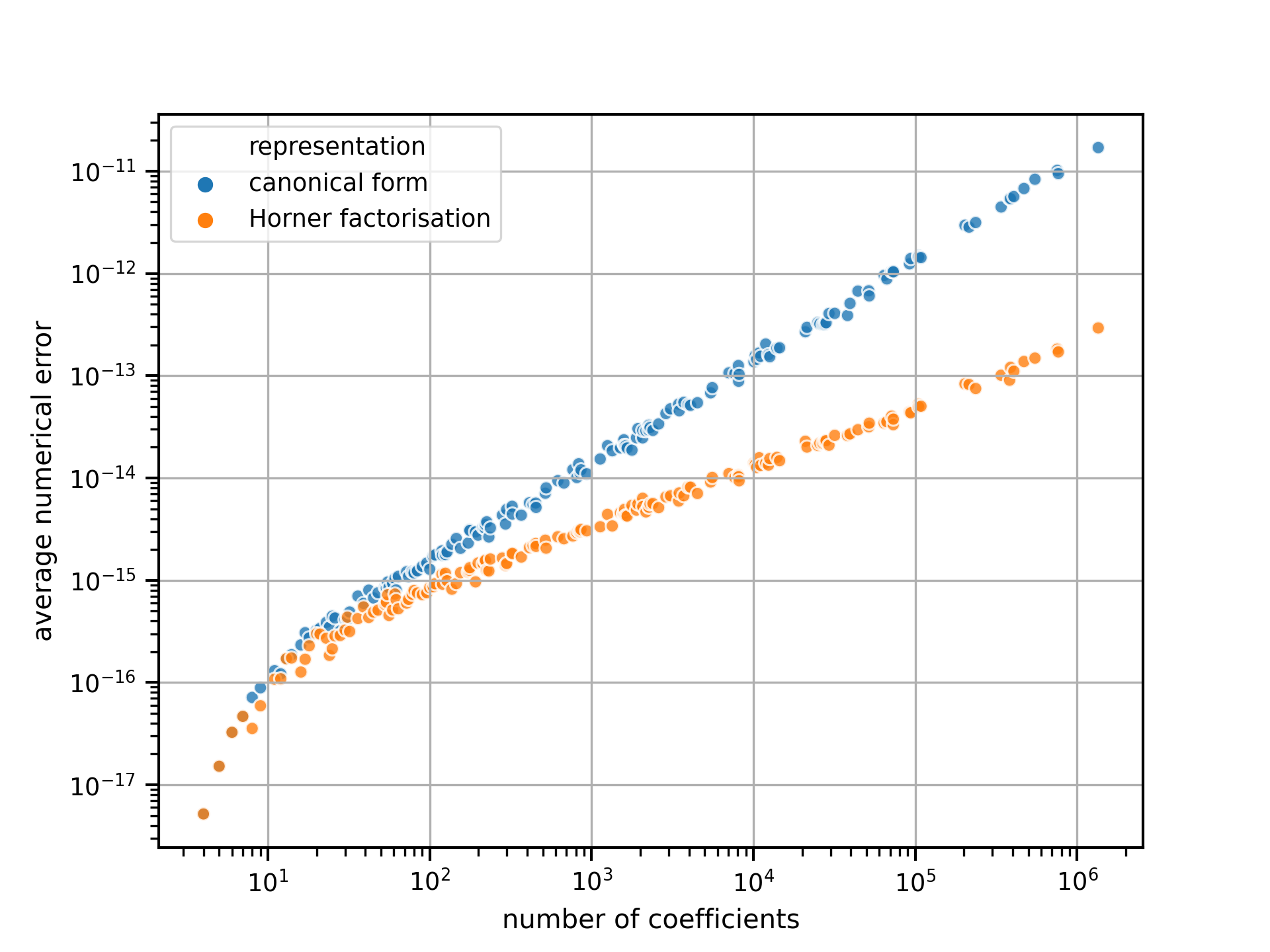}
\caption{numerical error of evaluating randomly generated polynomials of varying sizes.}
\label{fig:num_err_growth}
\end{figure}
\noindent

For benchmarking our method the following procedure is used:
In order to draw polynomials with uniformly random occupancy, the probability of monomials being present is picked randomly.
For a fixed maximal degree $n$ in $m$ dimensions there are $(n+1)^m$ possible exponent vectors corresponding to monomials.
Each of these monomials is being activated with the chosen probability.

For each maximal degree up to 7 and until dimensionality 7, 5 polynomials were drawn randomly.
In order to compute the numerical error, each polynomial has been evaluated at the point of all ones.
The true result in this case should always be the sum of all coefficients.
Any deviation of the evaluation value from the sum of coefficients hence is numerical error.
In order to receive more representative results, the obtained numerical error is being averaged over 100 tries with uniformly random coefficients in the range $[-1; 1]$.

\begin{figure}[tbh]
\centering
\includegraphics[width=\FigWidth]{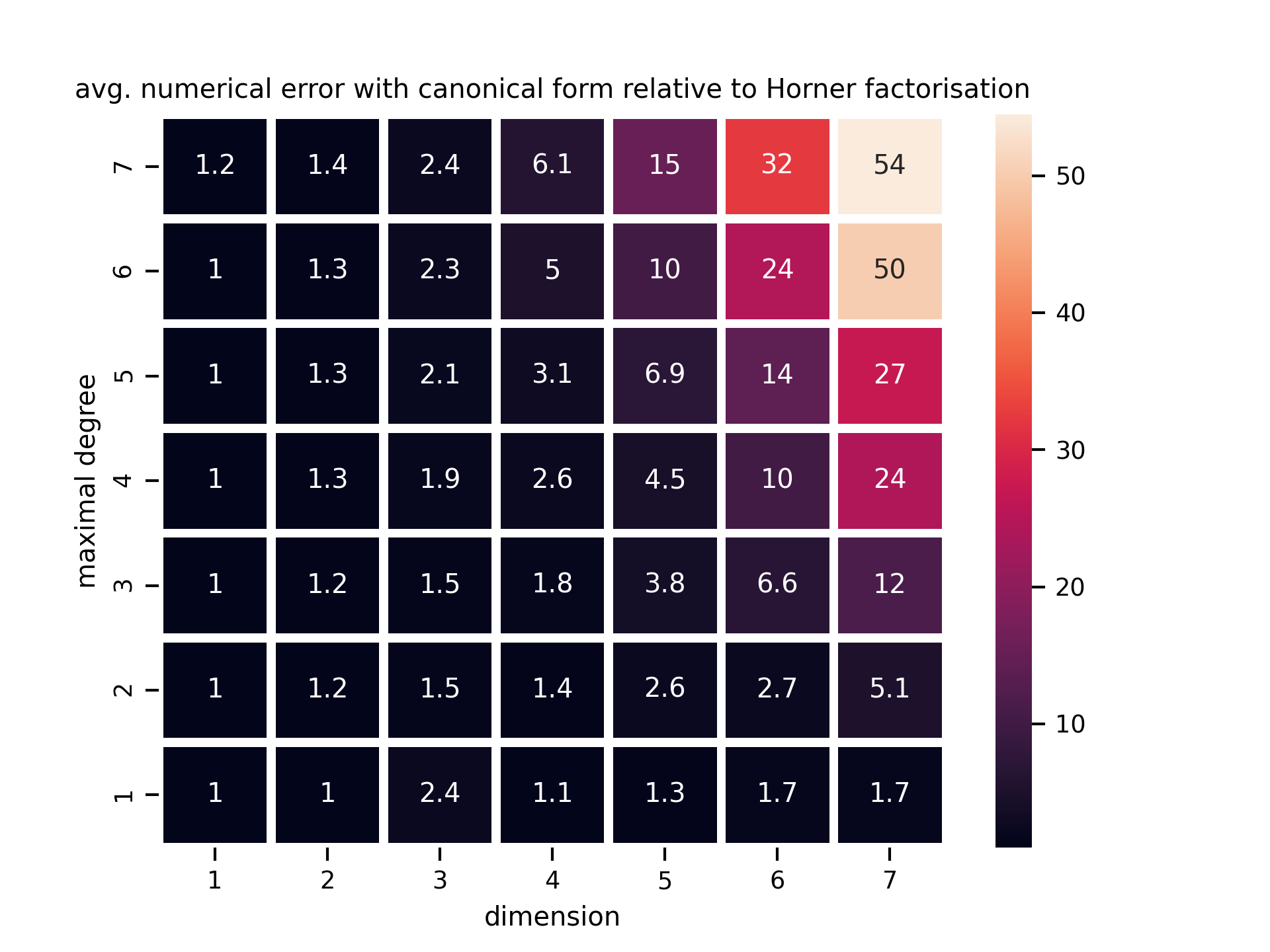}
\caption{numerical error of evaluating randomly generated polynomials in canonical form relative to the Horner factorisation.}
\label{fig:num_err_heatmap}
\end{figure}
\noindent

Note that even though the original monomials are not actually present in a Horner factorisation, the amount of coefficients however is identical to the amount of coefficients of its canonical form.
With increasing size in terms of the amount of included coefficients the numerical error of both the canonical form and the Horner factorisation found by \mbox{\texttt{multivar\_horner}} grow exponentially (cf. \cref{fig:num_err_growth}).
However, in comparison to the canonical form, the Horner factorisation is more numerically stable as it has also been visualised in \cref{fig:num_err_heatmap}.

\begin{figure}[tbh]
\centering
\includegraphics[width=\FigWidth]{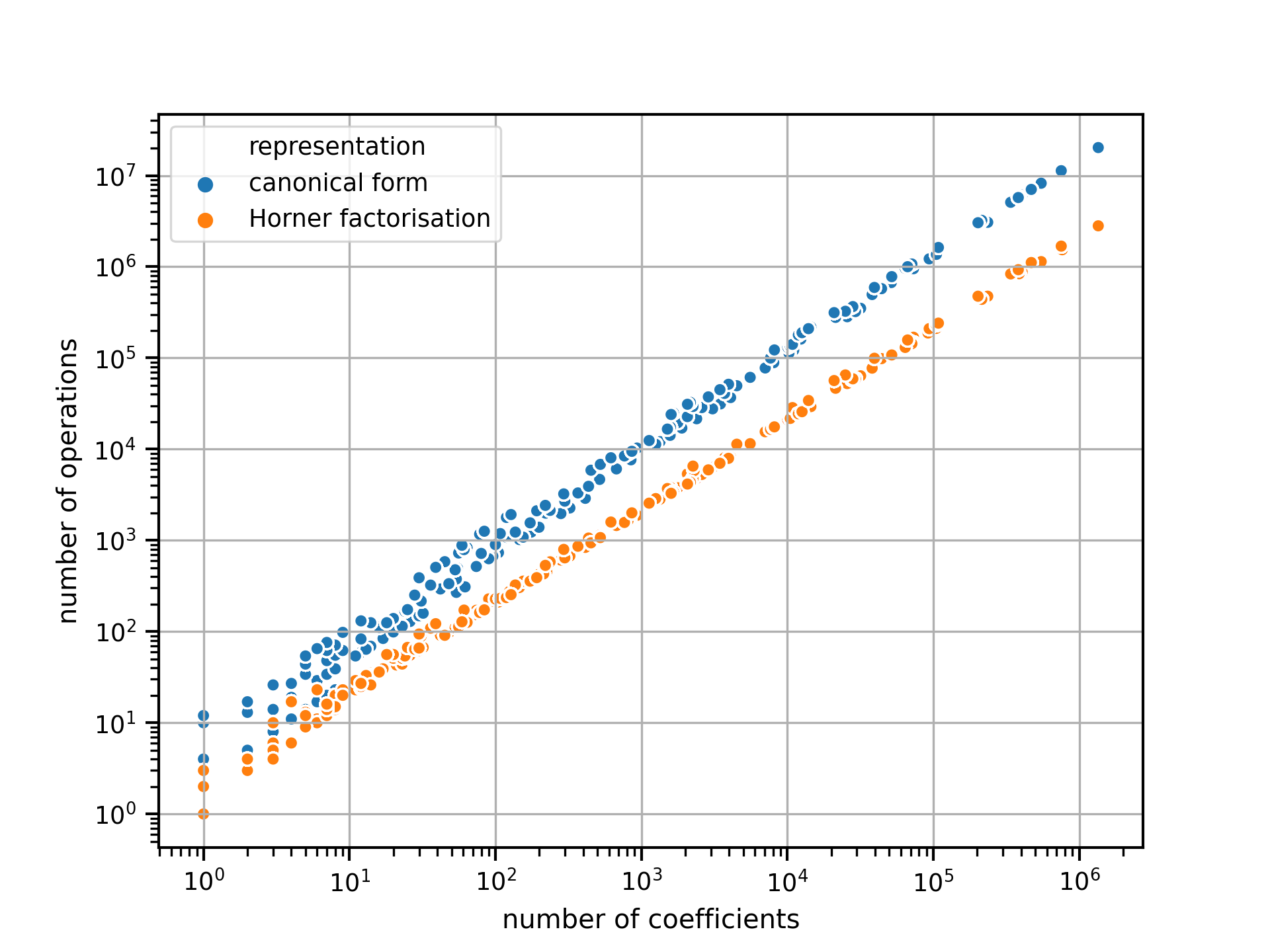}
\caption{amount of operations required to evaluate randomly generated polynomials.}
\label{fig:num_ops_growth}
\end{figure}

Even though the amount of operations required for evaluating the polynomials grow exponentially with their size irrespective of the representation, the rate of growth is lower for the Horner factorisation (cf. \cref{fig:num_ops_growth}).
As a result, the Horner factorisations are computationally easier to evaluate.

\noindent

\section{Related work}

The package has been created due to the recent advances in multivariate polynomial interpolation\cite{Hecht1, Hecht2}.
High dimensional interpolants of large degrees create the demand for evaluating multivariate polynomials computationally efficient and numerically stable.
Among others, these advances enable modeling the behaviour of (physical) systems with polynomials.
Obtaining an analytical, multidimensional and nonlinear representation of a system opens up many possibilities.
With so called \emph{interpolation response surface methods}\cite{michelfeitresponse} for example a system can be analysed and optimised.

The commercial software \href{https://www.maplesoft.com/support/help/Maple/view.aspx?path=convert%2Fhorner}{\mbox{\texttt{Maple}}} offers the ability to compute multivariate Horner factorisations. 
\mbox{\texttt{multivar\_horner}} however is the first open source implementation of a multivariate Horner scheme.
The \mbox{\texttt{Wolfram Mathematica}} framework supports \href{https://reference.wolfram.com/language/ref/HornerForm.html}{univariate Horner factorisations}.
The \mbox{\texttt{Julia}} package \href{https://github.com/JuliaAlgebra/StaticPolynomials.jl}{\texttt{StaticPolynomials}} has a functionality similar to \mbox{\texttt{multivar\_horner}}, but does not support computing Horner factorisations.

\href{https://numpy.org/doc/stable/reference/routines.polynomials.polynomial.html}{\mbox{\texttt{NumPy}}}\cite{numpy} offers functionality to represent and manipulate polynomials of dimensionality up to 3.
\mbox{\texttt{SymPy}} offers the dedicated module \href{https://docs.sympy.org/latest/modules/polys/index.html}{\mbox{\texttt{sympy.polys}}} for symbolically operating with polynomials.
As stated in the \href{https://mattpap.github.io/masters-thesis/html/src/algorithms.html#evaluation-of-polynomials}{documentation}, \mbox{\texttt{SymPy}} does not use Horner factorisations for polynomial evaluation in the multivariate case.
\href{https://doc.sagemath.org/html/en/reference/polynomial_rings/index.html}{\texttt{Sage}} covers the algebraic side of polynomials.

\mbox{\texttt{multivar\_horner}} has no functions to directly interoperate with other software packages.
The generality of the required input parameters (coefficients and exponents) however still ensures the compatibility with other approaches.
It is for example easy to manipulate a polynomial with other libraries and then compute the Horner factorisation representation of the resulting output polynomial with \mbox{\texttt{multivar\_horner}} afterwards, by simply transferring coefficients and exponents.
Some intermediary operations to convert the parameters into the required format might be necessary.

\section{Further reading}

The documentation of the package is hosted on \href{https://multivar_horner.readthedocs.io/en/latest/}{\texttt{readthedocs.io}}.
Any bugs or feature requests can be issued on \href{https://github.com/MrMinimal64/multivar_horner/issues}{\mbox{\texttt{GitHub}}}.
The \href{https://github.com/MrMinimal64/multivar_horner/blob/master/CONTRIBUTING.rst}{contribution guidelines} can be found there as well.

The underlying basic mathematical concepts are being explained in numerical analysis text books like \cite{neumaier2001introduction}.
The Horner scheme at the core of \mbox{\texttt{multivar\_horner}} has been theoretically outlined in \cite{greedyHorner}.

Instead of using a heuristic to choose the next factor, one can allow a search over all possible Horner factorisations in order to arrive at a minimal factorisation.
The amount of possible factorisations, however, is increasing exponentially with the degree and dimensionality of a polynomial (the amount of monomials).
One possibility to avoid computing each factorisation is to employ a version of $A^{*}$ search\cite{hart1968formal} adapted for factorisation trees.
\mbox{\texttt{multivar\_horner}} also implements this approach, which is similar to the branch-and-bound method suggested in \cite[ch. 3.1]{kojima2008efficient}.

\cite{carnicer1990evaluation} shows how factorisation trees can be used to evaluate multivariate polynomials and their derivatives.
In \cite{kuipers2013improving} Monte Carlo tree search has been used to find more performant factorisations than with greedy heuristics.
Other beneficial representations of polynomials are for example being specified in \cite{LeeFactorization2013} and \cite{leiserson2010efficient}.

\section{Acknowledgements}

Thanks to Michael Hecht\footnote{Max Planck Institute of Molecular Cell Biology and Genetics} and Steve Schmerler\footnote{Helmholtz-Zentrum Dresden-Rossendorf} for valuable input enabling this publication.

\printbibliography
\end{document}